%% file: Kitaev.tex
\documentclass[aps,prx,superscriptaddress,twocolumn]{revtex4-2}
\usepackage[utf8]{inputenc}
\usepackage{bm}
\usepackage{bbm}
\usepackage{bbold}
\usepackage{amsfonts}
\usepackage{amsmath}
\usepackage{amssymb}
\usepackage{mathtools}
\usepackage{upgreek}
\usepackage{xfrac}
\usepackage{soul}
\usepackage{xcolor}
\usepackage{comment}

\usepackage[colorlinks=true,citecolor=blue]{hyperref}
\hypersetup{colorlinks=true,citecolor=blue,linkcolor=blue ,urlcolor=blue}
\usepackage{url}

\usepackage{graphicx}
\usepackage{dcolumn}
\usepackage{physics}
\usepackage{tikz}
\usetikzlibrary{quantikz}
\usepackage{circuitikz}
\usepackage{dsfont}

\begin{document}

\title{Quantum subspace expansion approach for simulating dynamical response functions of Kitaev spin liquids}

\author{Chukwudubem Umeano}
\affiliation{Department of Physics and Astronomy, University of Exeter, Stocker Road, Exeter EX4 4QL, United Kingdom}
\author{François Jamet}
\affiliation{National Physical Laboratory, Teddington, TW11 0LW, United Kingdom}
\affiliation{IQM France SAS, 40 Rue du Louvre, 75001 Paris, France}
\author{Lachlan P Lindoy}
\affiliation{National Physical Laboratory, Teddington, TW11 0LW, United Kingdom}
\author{Ivan Rungger}
\affiliation{National Physical Laboratory, Teddington, TW11 0LW, United Kingdom}
\affiliation{Department of Computer Science, Royal Holloway, University of London, Egham, TW20 0EX, United Kingdom.}
\author{Oleksandr Kyriienko}
\affiliation{Department of Physics and Astronomy, University of Exeter, Stocker Road, Exeter EX4 4QL, United Kingdom}

\date{\today}

\begin{abstract}
We develop a quantum simulation-based approach for studying properties of strongly correlated magnetic materials at increasing scale. We consider a paradigmatic example of a quantum spin liquid (QSL) state hosted by the honeycomb Kitaev model, and use a trainable symmetry-guided ansatz for preparing its ground state. Applying the tools of quantum subspace expansion (QSE), Hamiltonian operator approximation, and overlap measurements, we simulate the QSL at zero temperature and finite magnetic field, thus moving outside of the symmetric subspace. Next, we implement a protocol for quantum subspace expansion-based measurement of spin-spin correlation functions. Finally, we perform QSE-based simulation of the dynamical structure factor obtained from Green's functions of the finite field Kitaev model. Our results show that quantum simulators offer an insight to quasiparticle properties of strongly correlated magnets and can become a valuable tool for studying material science.
\end{abstract}

\maketitle

\section{Introduction}

Studying materials with strong correlations is essential for discovering new phases of matter \cite{Morosan2012}, which in turn can enable further technological advances \cite{Kim2011,Yuan2023,Tabib-Azar2023}. Strongly correlated materials correspond to quantum systems where complex many-body interactions prevent describing them in the mean-field approach, and thus are difficult to study due to complexity increasing with system size \cite{Balents2010}. One example corresponds to the family of strongly correlated magnetic models \cite{Norman2016}, which includes quantum spin liquids \cite{Savary2017} and Kitaev materials \cite{Jackeli2009,Rau2014,Trebst2022rev,Rousochatzakis2024rev}. These systems can be described by two-dimensional (2D) models that exhibit long-range correlations \cite{Yang2008}, ground state entanglement \cite{Feng2022}, fractionalized statistics of excitations \cite{Takagi2019rev,Rousochatzakis2019}, among many other features \cite{Hermanns2018rev}. State-of-the-art numerical methods include density matrix renormalization group (DMRG) \cite{SCHOLLWOCK2011} with other tensor network (TN) approaches \cite{Cirac2009,Evenbly2015}, and quantum Monte Carlo (QMC) \cite{Suzuki1986,Blunt-Booth2015}. DMRG is instrumental for describing large systems with up to a hundred sites, but truncates correlations and typically assumes cylindrical topology \cite{Verstraete2023,Savary2017}. TNs offer excellent results for 1D systems, but tensor contractions for 2D models remain a challenge \cite{Haferkamp2020,Cirac2021rev}. QMC works successfully for lattices of varying dimensionality \cite{Sandvik1991,Balents2010} but is confined to non-zero temperature and can suffer from the sign problem \cite{Hangleiter2020}. All the aforementioned methods are suitable for studies of ground state properties, while dynamical studies remain a challenge. When studying zero temperature behavior in the absence of symmetries and with the need of evolving states, magnetic systems require using exact diagonalization (ED) \cite{Hickey2019}. For ED the state-of-the-art remains limited to $N=48$ sites \cite{Wietek2018,Wietek2024}. Consequently, the development of new numerical tools capable of simulating quantum dynamics on larger scales is necessary. To this end, an end-to-end quantum algorithmic toolbox supported by quantum simulation hardware with reduced noise is much desired by material scientists studying finite-size scaling of strongly correlated matter \cite{ALEXEEV2024}.

Quantum simulation (QS) and computing may offer a distinct powerful framework for studying materials at regimes that are inaccessible to classical numerics \cite{Bloch2012,Altman2021,Daley2023}. The goal of quantum simulation is to implement effective Hamiltonians of quantum systems with a controllable quantum hardware \cite{Henriet2020quantumcomputing,Ebadi2021,Daley2022}, and study their properties via state preparation \cite{Bosse-Montanaro2022,Michel2023}, unitary evolution \cite{Schmied2011,Kyriienko2018,Michel2024}, and measurements \cite{Kyriienko_inverse_alg,Cotler2019}. QS is particularly suitable for material science problems when hardware matches the lattice topology of the simulated system \cite{Houck2012,Li-Iadecola2023}, reducing the overheads for quantum runs. 
Towards the correlated matter studies, current state-of-the-art QS include Rydberg atom-based implementation for 2D topological spin liquid of Anderson type \cite{Semeghini2021}, Ising models with varying lattice geometry and ordering \cite{Labuhn2016,Scholl2021,Ebadi2021}, superconducting circuits-based QS of topologically ordered matter \cite{Satzinger2021}, discrete time crystals \cite{Mi2022,xiang2024longlivedtopo,bao2024schrodingercat} etc. Recent QS results on studying dynamics for the kicked transverse Ising model on a heavy-hex lattice \cite{Kim2023} suggested the quantum utility for such tasks run on 100+ qubit registers. And despite alternative classical solutions can still reach the scale \cite{Tindall2024,Begusic-Chan2024}, these simulations start to be very difficult to test \emph{in silico}. 
The advances of quantum hardware and reduced noise level thus open new opportunities for strongly correlated magnetism \cite{Rousochatzakis2024rev}. However, it is yet to be shown how the ability to evolve many-body Hamiltonians can be converted into relevant insights into material properties.
\begin{figure*}[t!]
\centering
\includegraphics[width=1.0\linewidth]{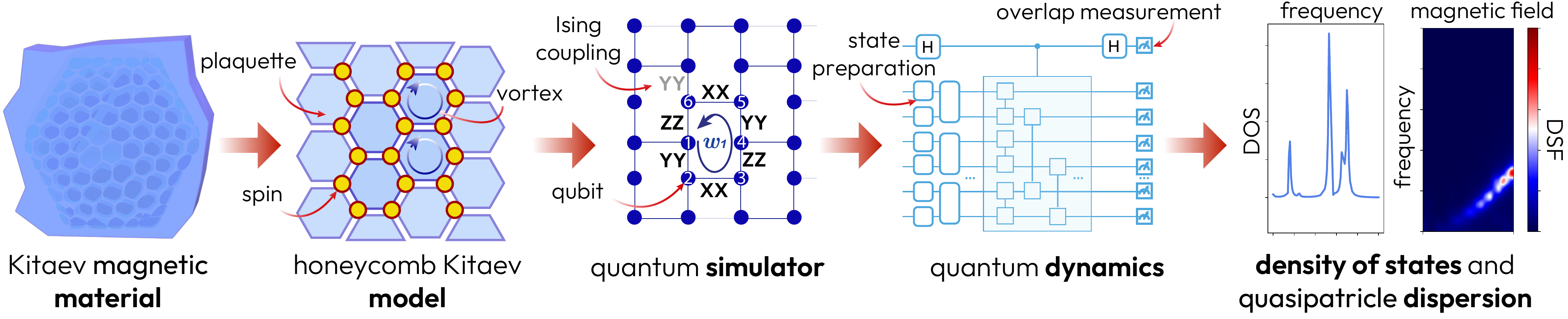}
\caption{Sketch of workflow for quantum subspace expansion-based approach for studying properties of Kitaev materials. First, material physics is mapped into a strongly-correlated spin model that can host a spin liquid (here represented by the honeycomb lattice Kitaev model for spin-1/2). Next, the spin Hamiltonian is mapped into a quantum simulator with the same coupling topology, representing a quantum digital twin. Then, we perform a state preparation and dynamical evolution with the system Hamiltonian, followed overlap measurements (via Hadamard test, reference state comparison, or equivalent). Finally, the measured overlaps are post-processed to distill the information about off-diagonal Green's functions for spin-spin correlations, dynamical structure factor (DSF), and quasiparticle response functions.}
\label{fig:Kitaev Model}
\end{figure*}

On the quantum software side, quantum simulation tasks can be split into two parts: ground state preparation \cite{Clinton2024} and simulation of dynamical properties \cite{Clinton2021}. Ground state preparation involves finding the lowest energy state of a quantum system, which can be achieved using the variational quantum eigensolver (VQE) for near-term devices \cite{Tilly20221rev,Cerezo2022rev}. VQE has been used to study molecular systems \cite{Grimsley2019,Fujii2022,Cao2022,Chan2024,Shang2023} and models of magnetic materials \cite{Li-Iadecola2023,Lyu2023symmetryenhanced,Cheng2023,Park2024hamiltonian,getelina2024qse}. 
A distinct approach for these tasks is to use dynamics and overlap measurements with quantum subspace expansion (QSE) \cite{Parrish-McMahon2019QFD,Stair2020,Cortes_krylov,Kyriienko_inverse_alg,bespalova2021hamiltonian,Urbanek2020,Klymko2022,motta2023subspacemethod,getelina2024qse}, also known as quantum filter diagonalization \cite{Parrish-McMahon2019QFD,bespalova2021hamiltonian}, where evolution-based ansatz allows studying low-energy physics of simulated models. This is closely related to Krylov-type methods \cite{Balzer2012,nandy2024quantumdynamicskrylovspace,kirby2024analysisquantumkrylovalgorithms} and their quantum equivalents \cite{Cortes_krylov,Shen2023realtimekrylov}. QSE avoids pitfalls of barren plateaus \cite{mcclean2018barren,larocca2024reviewbarrenplateausvariational}, and can be employed to access beyond-ground state physics \cite{Parrish-McMahon2019QFD}. Applications of quantum subspace expansion include studying electronic structure \cite{motta2023subspacemethodselectronicstructure}, calculating Green's functions via QS \cite{jamet2021krylov,QSE_GF,dhawan2023quantum,GreeneDiniz2024quantumcomputed,Gomes2023}, and estimating other response functions. This approach provides a powerful tool for probing the properties of strongly correlated models that are challenging to study using other methods \cite{ALEXEEV2024}.

In this work, we develop a quantum simulation-based workflow for studying properties of strongly correlated magnetic materials at increasing scale. Taking the honeycomb Kitaev model as an intriguing example \cite{Jahin-Iadecola2022,Kitaev_model_VQE}, we describe steps of preparing a quantum spin liquid state as an approximate ground state in the symmetric sector (Sec.~\ref{sec:Kitaev}), and effectively cooling the system to the symmetry-broken QSL via quantum subspace expansion (Sec.~\ref{QSE_GSP}). We show that already using few steps of evolution with nearest-neighbour Hamiltonian, supplemented with overlap measurements, we are able to reach QSL physics at zero-temperature. We design the measurement strategy that by using the same ingredients allows measuring dynamical response functions, density of states, and generic Green's functions for spin-spin correlations (Sec.~\ref{sec:GF}). From the selected measurements of Green's functions we map out the dynamical structure factor for the finite field Kitaev model (Sec.~\ref{sec:results}). Our work paves a road towards establishing a quantum simulation toolbox for future studies of quasiparticle properties of strongly correlated materials. 


\section{Kitaev model}
\label{sec:Kitaev}

We study the honeycomb Kitaev model --- a model of interacting spin-1/2s with exchange anisotropy on a honeycomb lattice, which can contribute to the description of compounds known as Kitaev materials (Fig.~\ref{fig:Kitaev Model}, left) \cite{Trebst2022rev}. The Kitaev model is composed of the Ising interaction terms with alternating basis, where different bonds correspond to XX, YY, and ZZ two-body interactions (Fig.~\ref{fig:Kitaev Model}, middle) \cite{Kitaev_2006,Savary2017,Hermanns2018rev}. The corresponding Hamiltonian for $N$ sites reads
\begin{equation}
\label{eq:H0}
    \hat{H}_0 = J^x\sum_{\langle ij \rangle\in\mathcal{X}}\hat{X}_i\hat{X}_j + J^y\sum_{\langle ij \rangle\in\mathcal{Y}}\hat{Y}_i\hat{Y}_j + J^z\sum_{\langle ij \rangle\in\mathcal{Z}}\hat{Z}_i\hat{Z}_j,
\end{equation}
where $\hat{X}_i,\hat{Y}_i,\hat{Z}_i$ are the Pauli operators acting on qubit $i$, and $J^\alpha$ $(\alpha = x,y,z)$ are Ising coupling constants. While general coupling can be considered, it is instructive to study an isotropic point of $J^\alpha \equiv J$, which leads to the competition between couplings. Here, $\mathcal{X}$, $\mathcal{Y}$ and $\mathcal{Z}$ are sets containing pairs of nearest neighbour sites (bonds). The sets of bonds can be defined to accommodate various boundary conditions. We consider the coupling arranged using periodic boundary conditions on a torus, since such configuration is most representative of thermodynamic limit physics \cite{Lahtinen2008,Kells2008}. The ground state of the Hamiltonian \eqref{eq:H0} hosts the quantum spin liquid phase of the Kitaev type, characterized by long-range correlations and interesting physics even at limited system sizes \cite{Hickey2019}. At the same time, it is known to be integrable in the absence of symmetry-breaking terms, usually analyzed in terms of Majorana fermions or vortex excitations \cite{Kitaev_2006,Kells2008}. For the extensive review of Kitaev physics we refer to specialized reviews in Refs. \cite{Savary2017,Trebst2022rev,Hermanns2018rev}, and studies of correlated materials including iridates ($\text{Na}_2\text{IrO}_3$) and ruthenates ($\alpha\text{-RuCl}_3$), among others \cite{Rousochatzakis2024rev}. 

Next, we introduce an additional term corresponding to the external magnetic field \cite{Rousochatzakis-Perkins2018,Hickey2019},
\begin{equation}
    \hat{H}_{\mathrm{magn}} = \sum_{i=1}^N h_i^x\hat{X}_i + \sum_{i=1}^N h_i^y\hat{Y}_i + \sum_{i=1}^N h_i^z\hat{Z}_i,
\end{equation}
where $h_i^\alpha$ are the Cartesian components of the magnetic field acting on site $i$. The full finite field Kitaev model Hamiltonian therefore is the sum of two terms, $\hat{H}=\hat{H}_0+\hat{H}_{\mathrm{magn}}$. Note that the major consequence of adding $\hat{H}_{\mathrm{magn}}$ to the pure honeycomb Kitaev Hamiltonian is breaking the symmetry corresponding to vortex conservation (to be discussed in the next subsection). This makes the model highly non-trivial to study, already at the level of quantum phase transitions in the ground state \cite{Hickey2019}. For instance, aspects of magnetic field dependence and the presence of a gapless spin liquid remain a matter of debate \cite{Zhang2022}.

Finally, we note that symmetry breaking can also be achieved by adding the next-nearest neighbor coupling terms and anisotropy. We stress that the developed quantum simulation approach is also suitable for studying generalized Kitaev models and Kitaev-Heisenberg
Hamiltonians that are closer to real materials \cite{Rousochatzakis2015,Banerjee2016,Rousochatzakis2024rev}, as long as some approximate initial state is prepared and QSE is performed. 


\subsection{Field-free ground state preparation}

Let us return back to the pure (zero-field) Kitaev model $\hat{H}_0$. This Hamiltonian is of particular interest as it is exactly solvable (meaning its observables can be studied with tractable `polynomial depth' calculations). Kitaev demonstrated this by reducing the model to free fermions in a static $\mathbb{Z}_2$ field \cite{Kitaev_2006}. While this method works well in the thermodynamic limit and studying dispersions, for finite size systems and states the treatment is less trivial and generally requires gauge fixing \cite{Lahtinen2008,Lahtinen2011}. If we desire to perform ground state preparation (GSP) for finite systems, keeping the spin-1/2 description, and remain in the symmetric subspace, one requires an alternative method. One option was proposed in Ref.~\cite{Kitaev_model_VQE} where a symmetry-aware ansatz for variational preparation was developed. Being in essence similar to geometric quantum machine learning \cite{Larocca2022PRXQ,JJMeyer2023PRXQ}, this has proven to be a trainable approach when symmetries are embedded to enable an equivariant workflow from initialization through variation, and measurement.
\begin{figure}[tb]
\centering
\includegraphics[width = 1.0\linewidth]{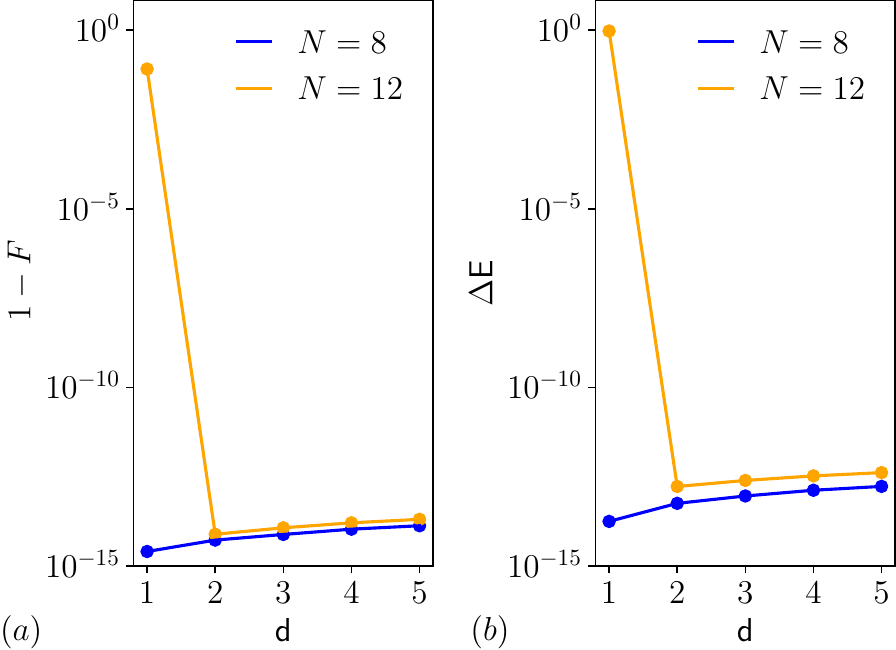}
\caption{Infidelity (a) and energy distance (b) between symmetry-aware variational ground state and exact Kitaev model GS, shown as a function of varying number of ansatz layers $d$. We consider $J=-1.0$, $h^z=0$ and $N=8$ and $N=12$. Variational ansatz was trained for 500 epochs and same performance is observed over various initialization of parameters (random uniform).}
\label{fig:VQE_plots}
\end{figure}

Let us briefly describe main ingredients for efficient GSP for the symmetric point of the Kitaev Hamiltonian, i.e. $\hat{H}_0$ \cite{Kitaev_model_VQE}. First, we need to define plaquette operators $\hat{w}_p$, which represent symmetry operators for the honeycomb Kitaev model. These operators are composed of the products of bond operators multiplied along a plaquette (loop). For instance, multiplying YY, XX, ZZ bond operators in the counter-clockwise direction, as shown in Fig.~\ref{fig:Kitaev Model}(middle), we obtain the plaquette operator $\hat{w}_1=\hat{X}_1\hat{Z}_2\hat{Y}_3\hat{X}_4\hat{Z}_5\hat{Y}_{6}$ (indices here are labelled simply by incrementing from the starting vertex; actual indexing depends on the system size considered). Working in the vortex basis, the expectation value of each $\hat{w}_p$ defines the presence or absence of a vortex in the corresponding plaquette $w_p$ \cite{Lahtinen2008,Savary2017}. Each plaquette operator commutes with all the others, as well as, crucially, the zero-field Kitaev Hamiltonian \cite{Hermanns2018rev},
\begin{equation}
    [\hat{w}_p,\hat{w}_{p'}] = 0 \enskip\forall~p,p', \quad [\hat{w}_p,\hat{H}_0] = 0 \enskip\forall~p.
\end{equation}
As a consequence, the number of vortices is conserved in the system. There are also two loop operators $\hat{l}_{x,y}$, defined as products of bond operators along a closed loop around the torus. These also define integrals of motion, as $[\hat{l}_{x,y}, \hat{H}_0] = 0$. Together with the plaquette operators, these define the stabilizer group \cite{LeeFlammia2017}
\begin{equation}
    \mathcal{S}=\{\hat{w}_p\}_{p=1}^n \cup \{\hat{l}_x, \hat{l}_y\}.
\end{equation}
Hence, these symmetries can be used to motivate the search for a suitable ansatz. The system is first initialized in a eigenstate of $\mathcal{S}$, before being put into the vortex eigenspace which contains the (degenerate) ground state. Since we wish to remain in the same vortex sector when we apply the parameterised ansatz $\hat{U}(\theta)$, we generate the circuit using a set of centralizers of the stabilizer group, $\mathcal{C(S)}$ \cite{Suchara2011}, which for this model are two-body Pauli operators acting along the bonds of the honeycomb lattice. Note that moving between sectors is easy, as they can be generated and annihilated in pairs by local rotations \cite{Kells2008}.

During VQE \cite{Tilly20221rev}, the ansatz's parameters are optimized to minimize the energy of the system, $E_\theta = \bra{\psi_0}\hat{U}(\theta)^\dagger\hat{H}\hat{U}(\theta)\ket{\psi_0}$, where $\ket{\psi_0}$ is the initial state prepared considering a selected symmetry sector. When the optimal parameters are found, $E_\theta \approx E_{\mathrm{GS}}$ and $\hat{U}_\theta\ket{\psi_0} \approx \ket{\psi_{\mathrm{GS}}}$. Employing the centralizer ansatz \cite{Kitaev_model_VQE} we show that the variational preparation of the Kitaev QSL ground state is efficient in the absence of magnetic field. The corresponding GSP runs are presented in Fig.~\ref{fig:VQE_plots}, showing that we achieved an energy distance $\Delta E = \abs{E_{\mathrm{GS}}-E_\theta} < 10^{-10}$ for all system sizes investigated, and similar results for the infidelity $(1-\abs{\bra{\psi_{\mathrm{GS}}}\ket{\psi_\theta}}^2)$. We note that this state preparation approach converges to the exact ground state at $d \sim N/6$, and the transition to the trainable regime is sharp. In fact, this linear depth requirement corresponds with the linear scaling of the Hamiltonian's dynamical lie algebra (DLA) via the overparameterization phenomenon \cite{larocca2023overparam}. The favorable overparametrization and small DLA thus enable high-quality GSP for $\hat{H}_0$.

In the presence of non-zero magnetic field the conservation laws for vortices are no longer valid. This makes symmetry-preserving ansatz inefficient. 
Although the ansatz can be adjusted to account for this symmetry breaking, the DLA scaling of the new Hamiltonian $\hat{H}$ is unfavourable; it was found that $d=50$ was required to prepare an approximate GS even at $N=8$ \cite{Kitaev_model_VQE}, on par with what is usually required for Hamiltonian variational GSP in the absence of extensive symmetries \cite{Bosse-Montanaro2022}. Hence, it is necessary to pursue alternative methods for Kitaev model GSP in the presence of symmetry-breaking terms, and magnetic field in particular.

\section{Quantum subspace expansion for ground state preparation}\label{QSE_GSP}

For the problem of ground state preparation of the Kitaev model with a magnetic field, we turn to quantum subspace expansion (QSE). This method has appeared under several different names within the field \cite{QFD_paper,QSE_GF,Quantum_krylov,Theory_QSD,Non-orth_VQE,klymko2021_evol,McClean_QSE}, but the idea is the same: represent the trial GS wavefunction as a linear combination of basis states. We begin by choosing some reference state $\ket{\phi_0}$. This state shall be easily preparable and, ideally, have a sizeable overlap with the true GS. A suitable choice in our case is the VQE-prepared GS for the zero-field $(h=0)$ Kitaev model.

We then build the GS basis by applying the time evolution operator $\hat{V}(t)=\exp(-it\hat{H})$ onto the reference state. We use the two-level multigrid time evolution approach introduced in Ref.~\cite{QSE_GF}, which is beneficial over standard methods as it reduces the circuit depth required to propagate to long times. Hence we define our basis states as 
\begin{equation}\label{GS_basis}
    \ket{\phi_{kl}}=\hat{V}(k\Delta_t)\bigg(\hat{V}\big((n_k+1)\Delta_t\big)\bigg)^l\ket{\phi_0},
\end{equation}
and the corresponding GS wavefunction as 
\begin{equation}
\label{GS_wavefn}
    \ket{\mathrm{GS}} = \sum_{l=-n_l}^{n_l}\sum_{k=k_{\min}}^{k_{\max}}\phi_{lk}^{\mathrm{GS}}\ket{\phi_{lk}}.
\end{equation}
Here, $l$ and $k$ are integers used to index the basis states. The values of $k_{\min}$ and $k_{\max}$ depend on $l$: for $l>0$, $(k_{\min},k_{\max})=(0,n_k)$; for $l<0$, $(k_{\min},k_{\max})=(-n_k,0)$; and for $l=0$, $(k_{\min},k_{\max})=(-n_k,n_k)$. The integers $n_k$ and $n_l$ determine the size of the basis, $n_\phi = 2(n_l+1)(n_k+1)-1$. 
The timestep $\Delta_t$ has to be selected carefully to ensure that the QSE converges to accurate energies quickly. Following Ref.~\cite{QFD_paper}, we set $\Delta_t= 2\pi/\kappa$, where $\kappa$ is a spectrum normalization factor estimated heuristically using the Gershgorin circle theorem.

The final step is to calculate the coefficients $\phi_{lk}^{\mathrm{GS}}$ to find the ground state energy (GSE), $E_{\mathrm{GS}} = \bra{\mathrm{GS}}\hat{H}\ket{\mathrm{GS}}$. These coefficients are determined using the variational Rayleigh-Ritz procedure, which corresponds to solving the generalized eigenvalue problem \cite{QFD_paper}
\begin{equation}\label{eigenproblem}
    \mathbf{H}\boldsymbol{\phi}^{\mathrm{GS}}=E_{\mathrm{GS}}\mathbf{S}\boldsymbol{\phi}^{\mathrm{GS}}.
\end{equation}
For this, we compute the Hamiltonian matrix $\mathbf{H}$ with  matrix elements, $H_{ij,kl} = \bra{\phi_{ij}}\hat{H}\ket{\phi_{kl}}$, and overlap matrix $\mathbf{S}$ with elements, $S_{ij,kl} = \bra{\phi_{ij}}\ket{\phi_{kl}}$. These two types of matrix elements are qualitatively different. For pure overlaps of propagated states $S_{ij,kl}$ we can use the Hadamard \cite{kitaev1997quantum,QFD_paper} or SWAP test (ancilla-based or destructive) \cite{Garcia-Escartin2013,Paine2023}. To minimize qubit requirement we can alternatively use suitable reference state as detailed in \cite{Kyriienko_inverse_alg,bespalova2021hamiltonian,Cortes_krylov,QSE_GF}, replace indirect measurements by direct ones  \cite{Mitarai2019}, employ classical shadows of increasing power \cite{Elben2023,Wan2023}, and recent state-of-the-art in the form of low-depth phase-sensitive measurements \cite{Yang-Cirac2024}.

For the Hamiltonian overlaps of the type $H_{ij,kl}$ we need an extra effort, as together with unitaries the Hermitian (but not unitary) operators must be implemented. While previously it was mostly done via Pauli averaging \cite{QSE_GF,Peng2021variationalquantum,BoPeng2022rev}, we propose to use the Hamiltonian operator approximation \cite{bespalova2021hamiltonian,Seki2021}, which allows decomposing $\hat{H}$ into a sum of evolution operators. This can be naturally incorporated into QSE scheme avoiding major overheads and circuit repetitions. Finally, once all elements are measured to sufficient precision, Eq.~\eqref{eigenproblem} is solved classically using a standard eigenproblem solver to determine $E_{\mathrm{GS}}$ and its corresponding eigenvector $\boldsymbol{\phi}^{\mathrm{GS}}$. As with all generalized eigenvalue treatment care must be taken to avoid instabilities, usually handled by adding a small regularization term and choosing a suitable algorithm. 


\section{Quantum subspace expansion for measuring correlation functions}
\label{sec:GF}

Now we proceed from the ground state preparation to studying excitation on top of GS. Here, suggest a QSE-based approach towards simulating (and measuring) correlation functions for spin systems that require dynamical quantum simulation. A procedure is related to protocols for calculating Green's functions (GFs) in fermionic systems \cite{QSE_GF}, while having the advantage of avoiding fermions-to-qubits mapping. Following Ref.~\cite{QSE_GF}, fermionic GFs can be calculated in the frequency domain, the retarded Green's function for fermions reads
\begin{equation}
    G_{\alpha\beta}(z) = G^{>}_{\alpha\beta}(z) + G^{<}_{\alpha\beta}(z),
\end{equation} 
with the greater and lesser GFs defined as 
\begin{align}
\label{Greater}
    &G^{>}_{\alpha\beta}(z) = \bra{\mathrm{GS}}\hat{c}_\alpha(z-\hat{H})^{-1}\hat{c}_\beta^\dagger\ket{\mathrm{GS}},\\ 
    \label{Lesser}
    &G^{<}_{\alpha\beta}(z) = \bra{\mathrm{GS}}\hat{c}_\alpha^\dagger(z+\hat{H})^{-1}\hat{c}_\beta\ket{\mathrm{GS}}.
\end{align}
Here, $\hat{c}_\alpha^\dagger$ and $\hat{c}_\alpha$ are the creation and annihilation operators acting on spin $\alpha$, and $z$ is the energy. Transitioning to spin systems, we choose to work with $\hat{c}_\alpha \mapsto \hat{Z}_\alpha$ operators at different sites (qubits), therefore studying information spread over the lattice. We focus on the calculation of the $\alpha=\beta$ instances greater GF [Eq.~\eqref{Greater}]. We note that the algorithm for the lesser GF [Eq.~\eqref{Lesser}] is analogous. The generalization to off-diagonal elements is straightforward, $G_{\alpha\beta}=(G^+_{\alpha\beta}-G_{\alpha\alpha}-G_{\beta\beta})/2$, with $ G^{+}_{\alpha\beta}(z) = \bra{\mathrm{GS}}(\hat{c}_\alpha+\hat{c}_\beta)(z-\hat{H})^{-1}(\hat{c}_\alpha^\dagger+\hat{c}_\beta^\dagger)\ket{\mathrm{GS}}$. We now drop the index $\alpha$ to simplify the notation.

Similar to the ground state preparation, quantum subspace expansion for calculating correlation functions relies on building a basis on top of some reference state. In this case the reference state is $\ket{\psi_0}=\hat{c}^\dagger\ket{\mathrm{GS}}$, where $\ket{\mathrm{GS}}$ is our QSE-prepared GS for $\hat{H}$. Our basis states in this case are
\begin{equation}
    \ket{\psi_{lk}}=\hat{V}(k\Tilde{\Delta}_t)(\hat{V}((\Tilde{n}_k+1)\Tilde{\Delta}_t))^l\ket{\psi_0}.
\end{equation}
We use these basis states to build a set of orthogonal Krylov states $\{\ket{\chi_n}\}$, 
\begin{equation}
    \ket{\chi_n} = \sum_{l=-\Tilde{n}_l}^{\Tilde{n}_l}\sum_{k=\Tilde{k}_{\min}}^{\Tilde{k}_{\max}}\psi_{lk}^n\ket{\psi_{lk}}.
\end{equation}
The tildes are used to distinguish the parameters used here from the similar parameters used previous for the GS basis~\eqref{GS_basis} and wavefunction~\eqref{GS_wavefn}. The orthogonal set of Krylov states $\{\ket{\chi_n}\}$ spans the space corresponding to the non-orthogonal states $\ket{\chi_0},\hat{H}\ket{\chi_0},\hat{H}^2\ket{\chi_0},\ldots,\hat{H}^n\ket{\chi_0}$ \cite{QSE_GF}. In this Krylov space, the Hamiltonian is tridiagonal and the diagonal greater GF can be written as a continued fraction,
\begin{equation}\label{continued_frac}
    G^>(z)=\frac{1}{z-a_0-\frac{b_1^2}{z-a_1-\frac{b_2^2}{z-a_2\ldots}}}.
\end{equation}
To determine the coefficients $\{a_n\},\{b_n\}$ we have to calculate the coefficients of the Krylov states $\psi_{lk}^n$. To do this, we calculate elements of the Hamiltonian matrix $\mathbf{H}_\psi$, the overlap matrix $\mathbf{S}_\psi$ and the transition matrix $\mathbf{S}_{\psi,\hat{c}^\dagger\phi}$, given by
\begin{align}
(\mathbf{H}_\psi)_{ij,kl} &= \bra{\psi_{ij}}\hat{H}\ket{\psi_{kl}},\\
(\mathbf{S}_\psi)_{ij,kl} &= \bra{\psi_{ij}}\ket{\psi_{kl}},\\
(\mathbf{S}_{\psi,\hat{c}^\dagger\phi})_{ij,kl} &= \bra{\psi_{ij}}\hat{c}^\dagger\ket{\phi_{kl}}.
\end{align}
%
%
%
%
As before, these matrix elements can be computed on a quantum computer with detailed techniques for measuring overlaps. By defining the first Krylov state $\ket{\chi_0} = \ket{\psi_0}$, we can compute the first set of coefficients $\boldsymbol{\psi}^0$ by
\begin{equation}
    \boldsymbol{\psi}^0 = \mathbf{S}_\psi^{-1}\mathbf{S}_{\psi,\hat{c}^\dagger\phi}\boldsymbol{\phi}^{\mathrm{GS}},
\end{equation}
with $\boldsymbol{\phi}^{\mathrm{GS}}$ being the set of coefficients of the QSE-prepared GS. The rest of the coefficients are determined iteratively as
\begin{align}
a_n &=\boldsymbol{\psi}^{n\dagger}\mathbf{H}_\psi\boldsymbol{\psi}^n,\\
b^2_n &=\boldsymbol{\psi}^{n-1\dagger}\mathbf{H}_\psi\mathbf{S}_\psi^{-1}\mathbf{H}_\psi\boldsymbol{\psi}^{n-1}-a_{n-1}^2-b_{n-1}^2,\\
\boldsymbol{\psi}^n &=\frac{1}{b_n}\left((\mathbf{S}_\psi^{-1}\mathbf{H}_\psi-a_{n-1})\boldsymbol{\psi}^{n-1}-b_{n-1}\boldsymbol{\psi}^{n-2})\right),
\end{align}
with $b_0=0$.
%
%
These coefficients are calculated on a classical computer, and inserted into Eq. \eqref{continued_frac} to calculate the GF. The iterative process can be terminated when one of the $b_n^2\approx 0$, as past this point higher order terms are insignificant.  
\begin{figure}[t!]
\centering
\includegraphics[width = \columnwidth]{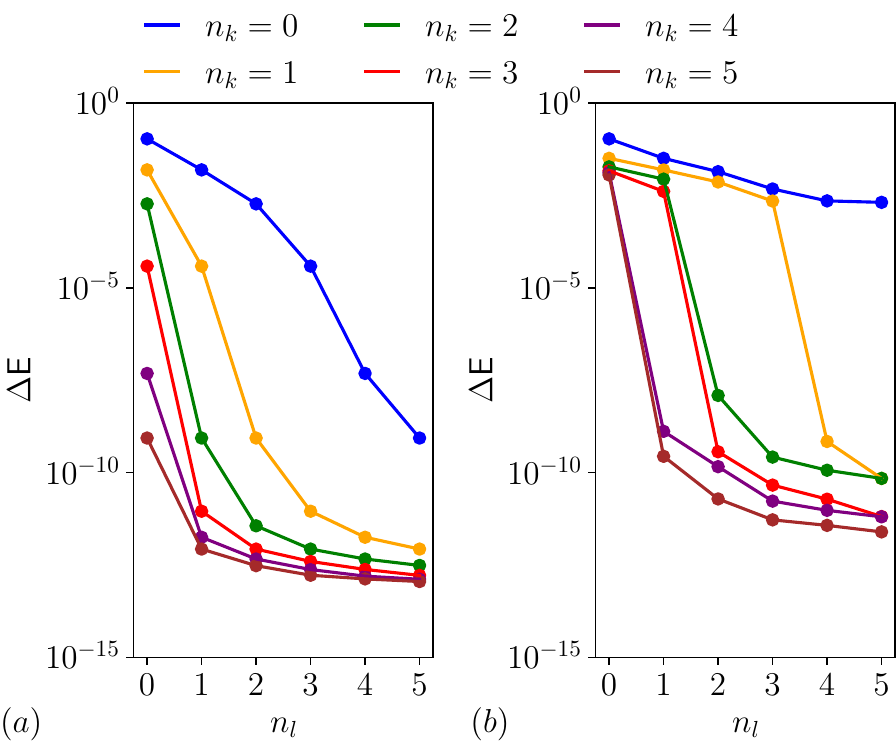}
\caption{Energy distance between GS prepared by QSE and exact Kitaev model GS for $N=8$, $J=-1.0$, $h^z=0.1$, as a function of $n_l$, for QSE with exact time evolution (a) and Trotterized time evolution with $r=1$ step (b).}
\label{fig:QSE_fix_n_k}
\end{figure}


\section{Results}
\label{sec:results}

\subsection{Implementation details}

We proceed with the numerical study the honeycomb Kitaev model for coupled spin-1/2s simulating the workflow of the quantum algorithm. As we need to assess the performance at high degree of precision, we perform statevector simulation as a starting point, and compare them with exact diagonalization-based results. For the torip periodic boundary conditions, we focus on systems with 4 and 6 plaquettes, corresponding to registers with $N=8$ and $N=12$ qubits, respectively. We consider the isotropic case, where $J^{x,y,z}=J$. When considering the model with a non-zero magnetic field, we choose a uniform magnetic field applied in the $z$ direction, so $\hat{H}_{\mathrm{magn}} = h^z\sum_{i=1}^N\hat{Z}_i$.

In Sec.~\ref{QSE_GSP} we mentioned that the reference state required for QSE is the variationally prepared GS for the field-free $(h^z=0)$ Kitaev model. Fig.~\ref{fig:VQE_plots} shows the results of the VQE procedure for $N=8$ and $N=12$ qubits. Here we plot the infidelity, $1-F = (1-\abs{\bra{\psi_{\mathrm{GS}}}\ket{\psi_\theta}}^2)$, and the energy distance, $\Delta E = \abs{E_{\mathrm{GS}}-E_\theta}$, between the variationally prepared GS and the exact GS. Here, $d$ denotes the number of layers of the symmetry-inspired ansatz used for VQE.

These results confirm that GSP for the field-free model using VQE works excellently. For $N=8$ just one layer is required to converge to the exact ground state up to numerical precision, while for $N=12$ only 2 layers are required. Using this information, we can select our reference states for QSE, as detailed in Table \ref{tab:VQE_details}.
\begin{table}[t!]
    \centering
    \begin{tabular}{|c|c|c|c|}
    \hline
        Model details & No. of layers $d$ & Infidelity & Energy distance  \\
        \hline
        $N=8,J=+1$ & 1 & $10^{-15}$ & $10^{-14}$ \\
        $N=8,J=-1$ & 1 & $10^{-15}$ & $10^{-14}$ \\
        $N=12,J=+1$ & 2 & $10^{-14}$ & $10^{-13}$ \\
        $N=12,J=-1$ & 2 & $10^{-14}$ & $10^{-13}$ \\
        \hline
    \end{tabular}
    \caption{Details of the VQE-prepared GSs used as the reference states for QSE. Infidelity and energy distance are measured between the variational GS and the corresponding exact GS.}
    \label{tab:VQE_details}
\end{table}

Throughout our investigation we make use of the Julia-based full state simulation package \texttt{Yao} \cite{Yao_paper}. As detailed previously, QSE relies on solving the generalized eigenvalue problem. We make use of the Locally Optimal Block Preconditioned Conjugate Gradient method (LOBPCG) \cite{LOBPCG_paper}, importing the solver from the \texttt{SciPy} python library.  


\subsection{Trotterization}

We rely on the application of the time evolution operator $\hat{V}(t)=e^{-it\hat{H}}$ with varying time steps to create the basis for QSE. On quantum circuit simulation packages such as \texttt{Yao}, we can apply $\hat{V}(t)$ as an exact operation as a reference for analog simulation mode. However, for digital simulators it is instructive to decompose the operator into a form which is implementable with finite gate sets available for certain platforms (e.g. Rydberg atoms or superconducting circuits). We note that several fast-forwarding methods exist for reducing the circuit depths for quantum simulation \cite{C_rstoiu_vff,kokcu2022cartan}, but the exponential scaling of the Kitaev Hamiltonian algebra makes these procedures challenging. Hence we opt for the second order Trotter-Suzuki approximation
\begin{equation}\label{Trotter_formula}
    e^{-i\hat{H}t} \approx \left(\prod_{s=1}^me^{\frac{-i\hat{H}_st}{2r}}\prod_{s=m}^1e^{\frac{-i\hat{H}_st}{2r}}\right)^r,
\end{equation}
where the Hamiltonian $\hat{H}=\sum_{s=1}^m\hat{H}_s$, and $r$ is the number of Trotter steps. Increasing $r$ reduces the Trotter error, at the expense of a linear increase in circuit depth. A balance has to be found to ensure that the algorithm works successfully while keeping circuit depths suitable for near-term quantum computation.
\begin{figure}[t!]
\centering
\includegraphics[width = 1.0\linewidth]{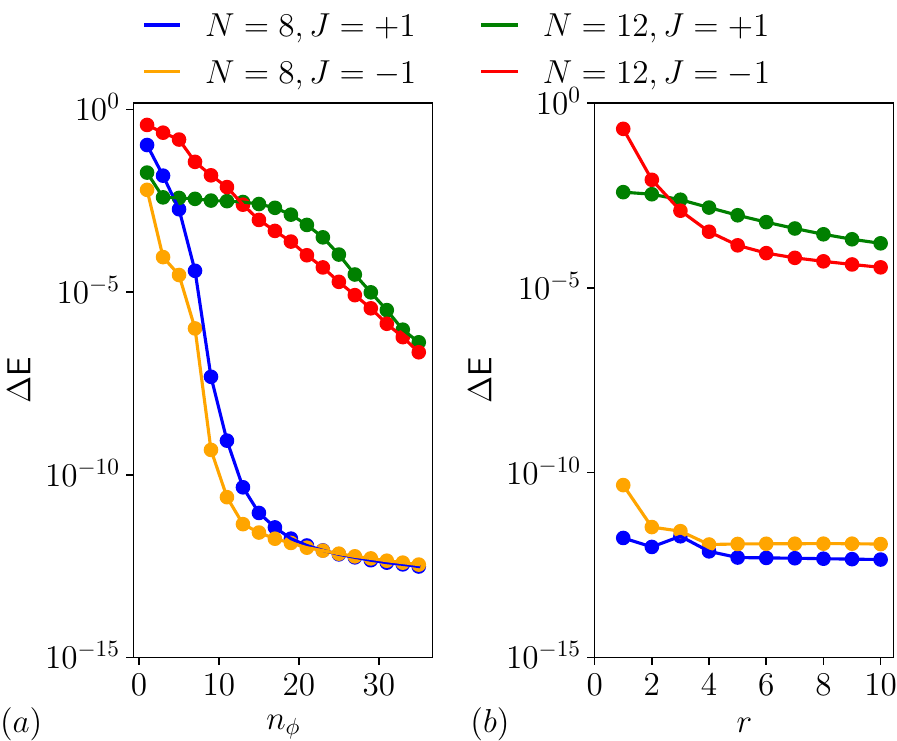}
\caption{(a) Energy distance between GS prepared by QSE with exact time evolution and exact Kitaev model GS as a function of number of basis states. (b) Energy distance between GS prepared by QSE with Trotterized time evolution and exact Kitaev model GS as a function of number of Trotter steps. For all calculations, the magnetic field was set as $h^z=0.1$. For the Trotterized QSE, we set $n_k=n_l=3$.}
\label{fig:QSE_basis_no_trot_steps}
\end{figure}

Each term $e^{-i\hat{H}_st}$ is a Pauli-string rotation, implementable using Pauli-Z rotations, CNOT gates, and basis rotations. The maximum circuit depth required to implement the two-level multigrid evolution (\ref{GS_basis}) is $d_v(n_l+1)$, where $d_v$ is the depth required to implement $\hat{V}(t)$ using Trotterization. Since the 2-qubit gates contribute the majority of the quantum circuit noise, we can measure the depth in terms of the number of layers of CNOT gates \cite{QSE_GF}.

In our case, all the 2-qubit Pauli-$X$ rotations in $\hat{H_0}$ act on different pairs of qubits, so all $N/2$ rotations can be applied in parallel within one layer. The same applies for the 2-qubit $Y$ and $Z$ rotations. Taking into account the symmetrization from the second order Trotter-Suzuki expansion (which also means that the two middle layers can be combined), as well as the number of Trotter steps $r$, we have $5r$ layers of rotations, with each rotation containing 2 CNOT gates. Hence $d_v = 10r$, and the total number of CNOT gates is $d_vN/2 = 5Nr$. 


\subsection{Kitaev model ground state preparation}

We now use quantum subspace expansion to effectively ``cool down'' the system into an approximate ground state of the full Hamiltonian $\hat{H}$. Fig.~\ref{fig:QSE_fix_n_k} shows results for the GSP using QSE with exact time evolution (a) and Trotterized time evolution (b) of the honeycomb Kitaev model. We calculate the energy distance $\Delta E$ for varying $n_l$ and $n_k$. As expected, increasing $n_l$ and $n_k$ significantly improves the QSE convergence to the exact GS for both exact and Trotterized time evolution, and we are able to reduce the error below $10^{-10}$ with a sufficient number of basis states. 
Furthermore, we can see that the error in QSE with exact time evolution only depends on the total number of basis states $n_\phi = 2(n_l+1)(n_k+1)-1$, rather than the individual values of $n_l$ and $n_k$. For example, as Fig.~\ref{fig:QSE_fix_n_k}(a) shows, the value of $\Delta E$ is virtually identical for $(n_l,n_k)$ = (0,5), (1,2), (2,1) and (5,0), which all correspond to $n_\phi = 11$. Similar behavior for dynamically-expanded GS was found in the case of quantum inverse iteration methods \cite{Kyriienko_inverse_alg}.

Following this, we can look at QSE convergence with increasing $n_\phi$, which we plot in Fig.~\ref{fig:QSE_basis_no_trot_steps}(a). Naturally, QSE is more difficult with larger system sizes, but we are still able to converge to the exact GS with errors $\leq10^{-6}$ for $N=12$ qubits with a large basis size.

The same applies in general for QSE with Trotterized evolution (Fig.~\ref{fig:QSE_fix_n_k}b), just that more basis states are required to achieve good convergence compared to QSE with exact time evolution. When using Trotterization, we have to be mindful of the distinct impacts of $n_k$ and $n_l$ on performance and resource requirements. An increase in $n_k$ causes an increase in the Trotter error for this evolution. However, as explained in \cite{QSE_GF,QFD_paper}, the QSE method is tolerant to these inaccuracies.
\begin{figure}[tb]
\centering
\includegraphics[width=1.0\linewidth]{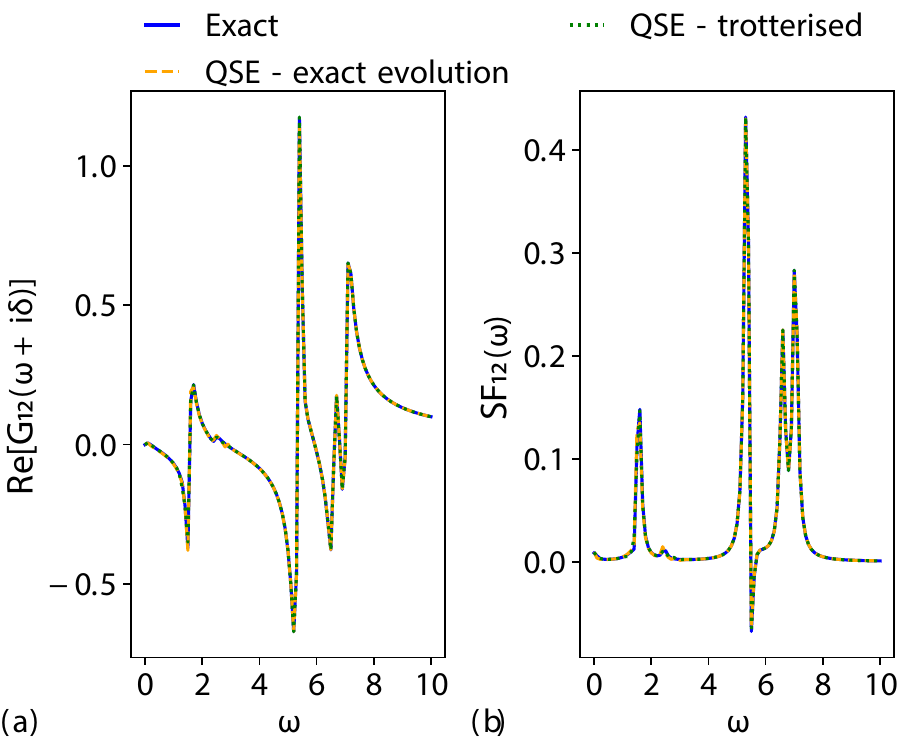}
\caption{Real part of the retarded Green's function $G_{12}$ (a) and the corresponding spectral function $\mathrm{SF}_{12}$ (b) of the Kitaev model for $N=8$, $J=-1$, $h^z=0.1$. We calculate both quantities on the spins $(1,2)$, i.e. $\hat{c}_\alpha = \hat{Z}_1$ and $\hat{c}^\dagger_\beta = \hat{Z}_2$. We use the same basis size as for the ground state preparation, i.e. $n_k=n_l=\Tilde{n}_k = \Tilde{n}_l=3$.}
\label{fig:GF_plots}
\end{figure}
By increasing $n_k$, we can increase the size of the basis $n_\phi$, as well as the maximum time reached, $T=((n_k+1)n_l+n_k)\Delta_t$, without increasing the total number of Trotter steps, $r(n_l+1)$. In summary, the two-level multigrid evolution helps us to reduce circuit depths for the same basis size, while maintaining high QSE performance.

We can also fix $n_k$ and $n_l$, and increase the number of (outer) Trotter steps $r$ used in the Trotter-Suzuki expansion (\ref{Trotter_formula}), as seen in Fig.~\ref{fig:QSE_basis_no_trot_steps}(b). The impact of reducing the Trotter error by increasing $r$ is particularly noticeable for the 12-qubit model.


\subsection{Kitaev model Green's function computation}

For the correlation function calculations for the honeycomb Kitaev model, we begin with the QSE-prepared GS with $n_k=n_l=3$. As mentioned, we used single qubit Pauli-$Z$ operators $\hat{Z}_\alpha$ as the creation/annihilation operators $\hat{c}^\dagger_\alpha/\hat{c}_\alpha$. For the GF basis we used the same variables as for the GS basis, so $\Tilde{\Delta}_t=2\pi/\kappa$ and $\Tilde{n}_k = \Tilde{n}_l=3$.

When considering Trotterized evolution, we use the same number of Trotter steps for the GS and GF basis, $r=\Tilde{r}=5$. We calculated the retarded Green's function for the off-diagonal elements of the form $G_{12}(z)$ for energies $z=\omega+i\delta$, where $\omega$ is a real energy and $\delta=0.1$. We do our calculations using QSE and compared the values to those calculated via exact diagonalization, as shown in Fig.~\ref{fig:GF_plots}(a).

We can see that when a sufficient basis size is used for both the GS basis and GF basis, our QSE calculations align almost perfectly with the exact diagonalization results. This is true even with the error introduced by Trotterization, demonstrating the robustness of the method to errors introduced by the imperfect basis creation. This bodes well for implementations of this method to study the Kitaev model on near term devices.
\begin{figure}[tb]
\centering
\includegraphics[width = 1.0\linewidth]{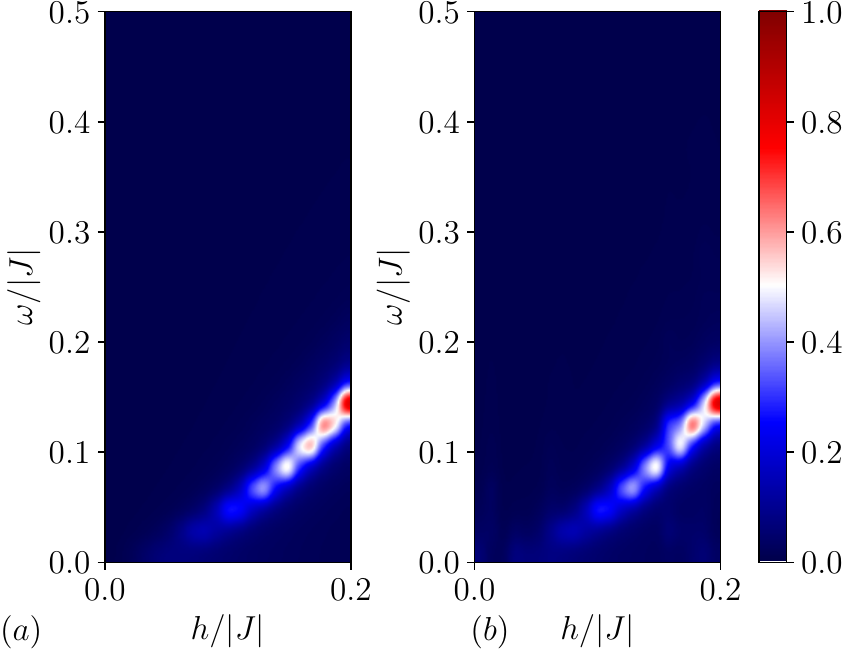}
\caption{Dynamical structure factor (DSF) $S_{\textbf{q}=0}(\omega)$ calculated using exact diagonalization (a), and quantum subspace expansion (b) approaches. Here, we considered the system with with four plaquettes, toric boundary conditions, and isotropic coupling of $J=-1$. The intensity values for DSF as normalized to fit in range [0,1].}
\label{fig:DSF_plots}
\end{figure}

From the GF, we can compute several other observables that may arise in studies within materials science. One of these is a spectral function (SF), which can be calculated directly from the off-diagonal retarded GF in the frequency domain: $\mathrm{SF}_{12}(\omega) \equiv -\frac{1}{\pi} \mathrm{Im} G_{12}(\omega+i\delta)$. We plot SF for the Kitaev model in Fig.~\ref{fig:GF_plots}(b).

We the extend the analysis further and calculate other correlation functions, such as the dynamical structure factor (DSF). This is given by \cite{Knolle_2015_DSF,Kaib_2019_DSF}
\begin{align}
    S_{\textbf{q}}(\omega)&=\sum_{\mu=x,y,z}S^{\mu\mu}_{\textbf{q}}(\omega)\\
    &=\sum_{\mu=x,y,z}\frac{1}{N}\sum_{ij}e^{-i\textbf{q}(\textbf{r}_i-\textbf{r}_j)}\int_{-\infty}^{\infty} dte^{i\omega t} S^{\mu\mu}_{ij}(t)
\end{align}
where $S^{\mu\mu}_{ij}(t)=\langle \hat{\sigma}_i^\mu(t)\hat{\sigma}_j^\mu(0) \rangle$. We can then rewrite the DSF in terms of the retarded GF,
\begin{equation}
    S_{\textbf{q}}(\omega)=\sum_{\mu=x,y,z}\frac{1}{N}\sum_{ij}e^{-i\textbf{q}(\textbf{r}_i-\textbf{r}_j)}\mathrm{Im}\big\{G^{\mu\mu}_{ij}(\omega)\big\}.
\end{equation}
We calculated the DSF with $\textbf{q}=0$, for a range of external magnetic field strengths and energies. The QSE calculations for the DSF used the same parameters as for the GF calculations: $n_k=n_l=3$ for the GS basis and $\Tilde{n}_k = \Tilde{n}_l=3$ for the Krylov basis. We compared to our QSE-computed DSF to exact diagonalization results in Fig.~\ref{fig:DSF_plots}, and we observe a clear qualitative agreement between the two plots. These correlation functions have a direct relation to observables measured within scattering experiments. Specifically, inelastic neutron scattering (INS) represents a well-known technique for studying material's DSF \cite{Banerjee2016}, making it a target for simulations in strongly correlated magnetism. Our toolbox shows that quantum simulators can contribute here as well. Therefore, our protocols facilitate the accurate studying of Kitaev model and spin liquid phenomena.\\


\section{Conclusions}

In this study we have demonstrated the application of the quantum subspace expansion to calculate high-fidelity ground states of the honeycomb Kitaev model with an external magnetic field. We have linked it with protocols for calculating correlation functions of the Kitaev model, employing quantum subspace expansion and Hamiltonian operator approximation. Using these calculations, we were able to extend to accurately compute other quantities of interest within condensed matter research, namely the spectral function as an imaginary part of the retarded Green's function, and the dynamical structure factor. The results from QSE strongly matched those produced by exact diagonalization, even with a limited number of Trotter steps used in the preparation of the ground state and Krylov bases.

Hence we have shown that QSE is a viable quantum simulation protocol for non-trivial strongly-correlated materials, with a resource budget suitable for implementations on near-term quantum hardware. It should be noted that studying the honeycomb Kitaev model is particularly pertinent due to the link to quantum error correction procedures, particularly toric codes. Beyond the Kitaev model, we envision that the QSE protocol can be extended to other complicated physical systems of interest within the material sciences community.


\begin{acknowledgements}
We acknowledge the discussions and useful suggestions for the manuscript by Kok Wee Song, Madhumita Sarkar, and Yannic Rath. We also thank our collaborators at Pasqal (Loïc Henriet and Constantin Dalyac) for discussions on quantum simulation approaches. The authors acknowledge the funding from UK EPSRC (award EP/Y005090/1).
\end{acknowledgements}


\input{Kitaev.bbl}
\end{document}

%% file: Kitaev.bbl
\providecommand{\noopsort}[1]{}\providecommand{\singleletter}[1]{#1}%